\documentclass[aps,amsfonts,nofootinbib]{revtex4-1}
\usepackage{epsfig}
\usepackage{graphicx}
\usepackage{amsmath}
\usepackage{amsbsy}
\def\openV{\leavevmode\hbox{ V \kern-9.8pt\normalsize V}}
\begin{document}
\newcommand{\ee}{\end{equation}}
\newcommand{\bb}{\begin{equation}}
\newcommand{\eqb}{\begin{eqnarray}}
\newcommand{\eqf}{\end{eqnarray}}
\def\sigmavec{\mbox{\boldmath$\sigma$}}
\def\x{\mathbf{x}}
\def\p{\mathbf{p}}
\def\ho{{\mbox{\tiny{HO}}}}
\def\sc{\scriptscriptstyle}
\newcommand{\1}{{\'{\i}}}
\def\sigmavec{\mbox{\boldmath$\sigma$}}
\def\nablavec{\mbox{\boldmath$\nabla$}}
\def\nc{{\mbox{\tiny{NC}}}}

\newcommand{\BS}{\boldsymbol}

\title{Aharonov-Bohm effect in a Class of Noncommutative Theories}
\author{Ashok Das}
  \email{das@pas.rochester.edu}
\affiliation{Department of Physics and Astronomy, University of Rochester, USA \\
and  Saha Institute of Nuclear Physics, 1/AF Bidhannagar, Calcutta-700 064, India}
\author{H. Falomir \email{falomir@gmail.com } and M. Nieto}
 \email{falomir@fisica.unlp.edu.ar }
\affiliation{IFLP/CONICET - Departamento de F\'{\i}sica, Facultad de Ciencias Exactas,
Universidad Nacional de La Plata, C.C. 67, (1900) La Plata, Argentina}
\author{J. Gamboa}
  \email{jgamboa55@gmail.com}
\affiliation{Departamento de  F\'{\i}sica, Universidad de  Santiago de
  Chile, Casilla 307, Santiago, Chile}
\author{F. M\'endez}
  \email{fernando.mendez.f@gmail.com}
\affiliation{Departamento de  F\'{\i}sica, Universidad de  Santiago de
  Chile, Casilla 307, Santiago, Chile}

\begin{abstract}
The Aharonov-Bohm effect including spin-noncommutative effects is considered.
At linear order in $\theta$, the magnetic field is gauge invariant although spatially strongly anisotropic. Despite this anisotropy, the Schr\"odinger-Pauli equation is separable
through successive unitary transformations and the exact solution is found. The scattering amplitude is calculated and compared with the usual case. In the noncommutative Aharonov-Bohm case the differential cross section is independent of $\theta$. 
\end{abstract}
\maketitle

\section{Introduction}

The discovery of noncommutative geometry has allowed the exploration of new directions in theoretical physics  \cite{1}.
For example, the noncommutative constructions of quantum field theory  \cite{2}, extensions beyond the relativistic
symmetry  \cite{3} and implications on condensed mater physics have been widely discussed \cite{4}.  This research has
also stimulated  the construction of new models in quantum mechanics \cite{ncqm}  which have a very natural transcription and --at
the same time--  has opened new windows and roads to explore (for example superconductivity \cite{super}).

In this context the algebra \cite{fglm} (we set $\hbar=1$)
\begin{align}
\left[{\hat x}_i,{\hat x}_j\right]  & = i\,\theta^2 \epsilon_{ijk} {\hat s}_k, &   \nonumber
\\
\left[{\hat     x}_i,{\hat    p}_j\right]    & =     i\,  \delta_{ij},
  &  & \left[{\hat    p}_i,{\hat p}_j\right ] = 0, \label{1}\\
\left[{\hat x}_i,{\hat  s}_j\right] & =  i\, \theta\, \epsilon_{ijk} {\hat
  s}_k, &  & \left[{\hat  s}_i,{\hat  s}_j\right] =  i \, \epsilon_{ijk} {\hat  s}_k,\nonumber
\end{align}
where $i,j, k = 1,2,3$  and $\theta$ is a  parameter with dimension
of  length, corresponds just to a deformation of the $\mbox{Heisenberg} \otimes
\mbox {spin}$ algebra which naturally induces spin-dipolar and higher order
interactions  as a result of  long-range spin interactions. The algebra (\ref{1}) is a non-relativistics version of the Snyder
algebra \cite{snyder}, which is rich enough to explore interesting consequences or
simplifications in real  physical systems.

Remarkably, the operators $(\hat{x}_i,\hat{p}_i)$ in (\ref{1}) can be realized in terms of the conventional dynamical variables satisfying the Heisenberg algebra of the usual coordinate and momentum operators $(x_i,p_i)$ through the relations
\eqb
 {\hat x}_i &=& x_i + \theta\, s_i, \nonumber\\
 {\hat   p}_i  &=&   p_i:=   -   \imath\, \partial_i, \label{2}\\
{\hat s}_i &=& s_i, \nonumber
\eqf
with
\eqb
\left[x_i,{ x}_j\right]  &=&0 = \left[p_i,{ p}_j\right],  \nonumber\\
\left[x_i, p_j\right]    &=&     i    \delta_{ij}, \label{3}
\eqf
where the matrices $s_i$, which commute with $(x_i,p_i)$, provide a $(2s+1)$-dimensional irreducible
representation (with $s$ integer or half-integer) of the $su(2)$ Lie algebra,
\bb
\left[{  s}_i,{  s}_j\right] =  i\,  \epsilon_{ijk} {  s}_k .
\nonumber
\ee
Therefore, with the realization in (\ref{2}), the Schr\"{o}dinger equation associated with the
Hamiltonian $H({\bf \hat{p}}, {\bf \hat{x}})$ can be written as
\bb
\imath\, \partial_t \psi (\mathbf{x},t)= H({\bf p}, {\bf x} + \theta\,{\bf s})\, \psi
(\mathbf{x},t) \,,
\label{9}
\ee
where $\psi (\mathbf{x},t)$ is a spinor with $(2s+1)$ components.

This simple observation, however,  has non-trivial and unexpected consequences as
can be seen by studying, for example,   the isotropic harmonic oscillator in this noncommutative space, given  by
the potential  $V({\bf \hat{x}})=\frac{ \omega^2}{2}\,{\bf \hat{x}}^2 $. Expressed in the basis of the normal
Heisenberg algebra \eqref{2}, it turns into
\eqb
V&=& \frac{\omega^2}{2} \left( {\bf x} +\theta {\bf s}\right)^2 \nonumber
\\
&=& \frac{1}{2} \omega^2 {\bf x}^2 + \frac{1}{2} \omega^2 \theta^2 {\bf s}^2 +
\omega^2 \theta {\bf x}\cdot{\bf s},
\eqf
where the term ${\bf x}\cdot{\bf s}$ becomes responsible --at least for particles with
spin $1/2$-- of the  infinite degeneracy of the ground state and the spontaneous
breaking of rotational symmetry, as shown in \cite{fglm}.

From these considerations as well as from the study of the effects induced by the algebra (\ref{1}) on other physical
systems one can speculate about the  order of magnitude of $\theta$.
For example, the present approach could be connected to recent experiments carried out using $^{52}$Cr condensates \cite{free,germ1,pfau1} where one uses the fact that, at large distances, the spins interact via the
potential
\bb
V =\mu  \left(\frac{{\bf s}_1\cdot{\bf  s}_2 -3({\bf s}_1\cdot  {\hat {\bf
      r}} )({\bf s}_2\cdot {\hat {\bf r}} )}{r^3}\right),
\label{dd}
\ee
where {${\bf  r}$ is  the relative position  vector, ${\hat  {\bf r}}={\bf r}/r$, and $\mu$
is the interaction strength (see {\it e.g.} \cite{landau1}). This potential can, in fact, be completely rederived from the present approach to noncommutative quantum mechanics with \cite
{nos}
\bb
\mu = \theta^2.
\ee

Following reference \cite{pfau1}, one can relate $\theta$ with the parameters of the experiment involving a gas of  $^{52}$Cr with total spin 3 to find
$$
\theta^2=\frac{C_{dd}}{4\pi}=\frac{48 a_0\hbar^2}{m},
$$
where  $a_0$ represents the  Bohr's radius  while $m$  is the mass of $^{52}$Cr isotope. This leads to the value \cite{nos}
\begin{equation}
\theta\sim 10^{-11} \mbox{cm},
\end{equation}
which would not change significantly
for spin-$1/2$ atoms, even though the experiments involving this kind of particles are, in principle,
much more complicated.

Taking into account these results, a natural question arises about the implications of algebra \eqref{1} on the topological nature of systems of physical interest. This is a non trivial question since the spin degrees of freedom now appear in the commutation relations \eqref{1} of the coordinates. The Aharonov-Bohm effect turns out to be paradigm in the study of such issues. Here one considers a charged particle in a  magnetic field which is
zero everywhere except inside an infinitely long and impenetrable solenoid of essentially zero radius. The dynamics, in this case, reduces to the motion in a plane perpendicular to the external magnetic field.

The goal of this paper is to generalize the study of such a system to a particle moving in a noncommutative space defined by \eqref{1}. We determine the Sch\"odinger-Pauli equation of a particle moving under the action of an external magnetic field, study its properties and evaluate the scattering amplitude. For simplicity, we will restrict ourselves to a plane orthogonal to the magnetic field
and consider the motion of a charged spin-$1/2$ particle. The calculations will be done to the leading order in $\theta$ in perturbation theory because of the mathematical difficulties posed by this problem. The paper is organized as follows. In Section {\bf II}, the gauge potential in the noncommutative space is discussed and in 
Section {\bf III} the Schr\"odinger-Pauli equation is
determined. In Section {\bf IV}, the scattering amplitude is evaluated and in Section {\bf V} we discuss briefly the physical implications of these results.

 \section{The Aharonov-Bohm Potential in the Noncommutative Plane} \label{NC-plane}

In this section we calculate the vector potential in the noncommutative plane and
discuss only those  physical properties which are useful in the study of the Aharonov-Bohm effect, with special attention to its topological  properties. Let us start with the  vector potential ${\bf A}$ in the
commutative case. Let us  denote the coordinates of the normal plane by $x_i, i=1,2$
and $r^2 =x_1^2 + x_2^2$. The vector potential which gives rise to a zero magnetic
field everywhere except at $r=0$  with finite flux is given by
 \bb \label{normalA}
 A_i = - \frac{\alpha}{2}\, \epsilon_{ij} \partial_j (\ln r^{2}) = -\alpha \epsilon_{ij} \frac{x_j}{r^2},
 \ee
with $\alpha$ a constant proportional to the magnetic flux $\Phi_B=2 \pi\alpha$. The
potential \eqref{normalA} is defined in the Coulomb gauge satisfying $\partial_i A_i=0$, and the
magnetic field along the $z$-axis has the form
 \begin{equation}
 B(r) = \epsilon_{ij} \partial_i A_j = -\alpha \epsilon_{ij}\partial_i \epsilon_{jk}
\partial_k \left( \ln r\right) = \alpha \mbox{\boldmath$\nabla$}^{2} \left( \ln r\right) = 2\pi \alpha \,\delta (r)= \Phi_B \,\delta (\mathbf{x}). \label{II1}
 \end{equation}

In the following we consider the corresponding potential in the noncommutative plane,
\begin{equation}
 A_i^{\nc}=-\frac{\alpha}{2}\,\epsilon_{ij} \partial_j \left(\ln {\hat{r}^2}\right)\,.
\end{equation}
The realization of the deformed algebra in (\ref{2}) leads to the replacement
 \bb
 \hat{r}^2 = r^2 +  \theta \,{\bf x}\cdot \mbox{\boldmath $\sigma$} + \frac{1}{2}\,\theta^2,
 \ee
where we have identified  ${\bf s} = \frac {1}{2} \mbox{\boldmath $\sigma$}$ with $\sigma_i\,, i=1,2$ denoting the two Pauli matrices.
Therefore the Cartesian components of the gauge potential in the noncommutative plane are defined as (matrices)
 \begin{equation}
 A_i^{\nc} =  -\frac{\alpha}{2} \epsilon_{ij} \partial_j \left(\ln \hat{r}^2\right) =  -\frac{\alpha}{2} \epsilon_{ij} \partial_j \left[ \ln \left(r^2 +\frac{\theta^2}{2}\right)
{\bf 1}_2 + \ln \left(  {\bf 1}_2 + \frac{2\theta \, {\bf x}\cdot \mbox{\boldmath $\sigma$}}{2r^2 +\theta^2}
\right)\right]. \label{II2}
 \end{equation}
It is possible to factor out all the spin dependence in the previous expression by defining
  \bb
  {\sigma}_{r} = \frac{{\bf x}\cdot\mbox{\boldmath $\sigma$}}{|{\bf x}|}= \mathbf{\hat{u}}_r\cdot\mbox{\boldmath $\sigma$}  =
  \left(\begin{array}{ll}0 &e^{-i \phi}
  \\
e^{i \phi}&0\end{array}
\right) , \label{II3}
  \ee
 and  its derivative
  \bb
\sigma_{\phi} :=\frac{\partial \sigma_{r}}{\partial \phi}= \mathbf{\hat{u}}_\phi \cdot \mbox{\boldmath $\sigma$}= i\,
\left(\begin{array}{ll}
0 &-e^{-i \phi}
\\
e^{i \phi}&0
\end{array}\right),
\label{sig2}
\ee
with $\phi$ denoting the polar angle. Here $\mathbf{u}_{r}, \mathbf{u}_{\phi}$ denote respectively the unit vectors along the radial and the azimuthal angle directions.These matrices obviously satisfy
 \bb
 \sigma^2_{r} = \mathbf{1}_{2} \label{sig1},~~~~~ \sigma^2_{\phi} = \mathbf{1}_{2},
 ~~~~~\sigma_{r}\sigma_{\phi}=i \, \sigma_3=
 - \sigma_{\phi}\sigma_{r}.
\ee

Using the Taylor series expansion for the logarithm, a straightforward calculation leads to
\begin{equation}\label{28}
    \begin{array}{c}
      \displaystyle
      A_i^{\nc}=A_i(r)\, f(\mathbf{x}, \theta)
    -\frac{\alpha}{4} \,\frac{x_i}{r^2}
    \,  \ln\left( \frac{{r^2+\left(r+\theta\right)^2}}{{r^2+\left(r-\theta\right)^2}}\right)
 \sigma_{\phi} \,,
    \end{array}
\end{equation}
where $A_i(r)$ is the vector potential in the commutative plane defined in Eq.\ (\ref{normalA}), and the function $f (\mathbf{x}, \theta)$ is given by
\begin{align}\label{29}
      f(\mathbf{x}, \theta) & =r \left\{\left( \frac{\left(2r+\theta \right)}{{r^2+\left(r+\theta\right)^2}}\right)
    \left( \frac{\mathbf{1}_2+{\sigma}_{r}}{2} \right)
    +\left( \frac{\left(2r-\theta \right)}{{r^2+\left(r-\theta\right)^2}}\right)
    \left( \frac{\mathbf{1}_2-{\sigma}_{r}}{2} \right)
    \right\}\nonumber\\
     & = \frac{4 r^4}{4 r^4+\theta^4}\, \mathbf{1}_2
       -r \theta \, \frac{2 r^2 -\theta^2}{4 r^4 + \theta^4}\, {\sigma}_{r}
       \stackrel{\sc \theta\rightarrow 0}{\longrightarrow} \mathbf{1}_2\,,
\end{align}
so that $A_{i}^{\nc}$ reduces to $A_{i}$ (as a diagonal matrix) when $\theta = 0$. On the other hand, for $\theta \neq 0$, $f(\mathbf{x}, \theta)$ is a continuous function of $\mathbf{x}$ vanishing at the origin.

We see from \eqref{28} that the vector potential in the noncommutative plane has a component in the
radial direction ($\mathbf{u}_{r}$) which is not present in the standard case. In fact, in polar coordinates we can write
 \bb
 {\bf A}^{\nc} = {\alpha} \left[
 \frac{4 r^3 \mathbf{1}_2- \theta(2 r^2 -\theta^2) {\sigma}_{r}}{4 r^4+\theta^4}
\right] \mathbf{\hat{u}}_\phi -\frac{\alpha}{4 r}  \ln\left(
\frac{2r^2 +2r\theta +\theta^2}{2r^2 -2r\theta +\theta^2}\right)\sigma_{\phi}\, {\mathbf{\hat{u}}_r}.  \label{GNC}
 \ee
For large $r$ with a finite $\theta$, this vector potential vanishes as
\begin{equation}\label{A-large-r}
    {\bf A}^{\nc} \longrightarrow \frac{\alpha}{r}  \left\{
   \left(\mathbf{1}_2-\frac{\theta}{2 r}\,
   {\sigma_r}\right){\mathbf{u}_\phi}-\frac{\theta}{2 r}\,
   \sigma_\phi  \, \mathbf{u}_r
   + O\left(\left(\frac{\theta}{r}\right)^3\right)\right\}\,,
\end{equation}
while near the origin we find a significant departure from Eq.\ (\ref{normalA}),
\begin{equation}\label{A-small-r}
    {\bf A}^{\nc} \stackrel{\sc r\rightarrow 0}{\longrightarrow} \frac{\alpha}{\theta}\left\{
   \left(1-\frac{2 r^2}{\theta^2}\right){\sigma_r}\, {\mathbf{u}_\phi}-
   \left(1-\frac{2 r^2}{3 \theta^2}\right)\sigma_\phi\, \mathbf{u}_r+
   O\left(\left(\frac{r}{\theta}\right)^3\right)\right\}\,.
\end{equation}

Some comments are in order here. First we note from its definition in Eq.\ (\ref{II2}) that this vector potential
satisfies the Coulomb gauge condition, $\partial_i A^{\nc}_i=0$ (to all orders in $\theta$).  In the noncommutative case, the magnetic field is  defined as the (gauge covariant) commutator of the covariant derivatives
\begin{equation}\label{covariant-derivatives}
    B^{\nc}=F_{12}^{\nc}=\frac{\imath}{e}\left[\partial_1-\imath e A^\nc_1 , \partial_2-\imath e A^\nc_2\right]=
    \partial_1  {A}^{\nc}_2 -\partial_2  {A}^{\nc}_1 - \imath e\left[{A}^{\nc}_1,{A}^{\nc}_2\right]\,,
\end{equation}
where the last term in the right hand side is non-vanishing since $[\sigma_r,{\sigma}_\phi]=2 \imath \sigma_3$. However, this leads to an $O\left({\theta}^2\right)$ contribution. In fact, a straightforward calculation leads to
\begin{align}\label{44}
      B_{(1)}^{\nc}(\mathbf{x}) & =\boldsymbol{\nabla} \times \mathbf{A}^{\nc}\nonumber\\
      & =\alpha \left(\frac{16 r^2 \theta^4}{\left(4 r^4+\theta^4 \right)^2}\right) \mathbf{1}_2+ \alpha\left\{\theta
    \frac{\left(2 r^2+
    \theta^2\right) \left(4 r^4-8 r^2\theta ^2+\theta^4\right)}{r \left(4r^4+\theta ^4\right)^2}
    +\frac{1}{4r^2}\,
      \ln\left( \frac{r^2+(r+\theta)^2}{r^2+(r-\theta)^2}\right)\right\} \mathbf{\sigma}_r
      \,,
\end{align}
while for the commutator in Eq.\ (\ref{covariant-derivatives})
\begin{equation}\label{F3-1}
    B_{(2)}^{\nc}(\mathbf{x})= - \imath e\left[{A}^{\nc}_1,{A}^{\nc}_2\right]
    =  \left( \frac{e}{c}\right) \frac{\alpha^2}{16 r} \,
       \partial_r \ln^2\left( \frac{{r^2+\left(r+\theta\right)^2}}{{r^2+\left(r-\theta\right)^2}}\right)
    {\sigma}_3\, ={\cal O} (\theta^2).
\end{equation}
It can be checked in a straightforward manner that the magnetic field $B^{\nc}(r,\phi)=B_{(1)}^{\nc}+B_{(2)}^{\nc}$ is invariant under rotations along the $z$-axis generated by
\begin{equation}\label{45}
    U(\gamma):= e^{\imath \gamma J} = e^{\imath \gamma \left(L+\frac{\sigma_3}{2} \right)}\,,
\end{equation}
where $J$ is the total angular momentum along the $z$-axis on the noncommutative plane while $L=\mathbf{x}\times\left(-\imath \boldsymbol{\nabla}\right)$ corresponds to the orbital angular momentum.

We note that to leading order in $\theta$ the magnetic field has been \emph{smeared out} by the non-commutativity of coordinates, resulting in a well behaved function near the origin (for $\theta \neq 0$),
\begin{equation}\label{smeared}
    B^{\nc}(r,\phi)=\frac{2 e \alpha }{\theta^2}\, {\sigma_3}-
    \frac{16 \alpha }{3\theta^3}\,r\, {\sigma_r}+
    \frac{16 \alpha }{3 \theta^4}\, r^2 \left(3\, \mathbf{1}_2 -
    e \, {\sigma_3}\right)+O\left(r^3\right).
\end{equation}
The flux of the magnetic field through a circle of radius $r$ can be easily evaluated.
 To first order in the non-commutativity parameter $\theta$, the vector potential in
Eq.\ (\ref{GNC}) gives rise to a magnetic field vanishing everywhere  except at the origin, with a magnetic flux as in the conventional  case,
 \bb
\int_{0}^{2\pi} r d\phi \mathbf{u}_{\phi}\cdot \mathbf{A}^{\nc} =  \int_0^{2\pi} r\, d\phi\,\mathbf{u}_\phi \cdot  \frac{\alpha}{r}  \left\{
   \left(\mathbf{1}_2-\frac{\theta}{2 r}\,
   {\sigma_r}\right){\mathbf{u}_\phi}-\frac{\theta}{2 r}\,
   \sigma_\phi  \, \mathbf{u}_r
   \right\}= 2 \pi \alpha \, \mathbf{1}_2\,,
 \ee
where we have used the fact that the integrals of $\sigma_r$ over the angular coordinate $\phi$ vanishes.
This suggests that, to linear order in the noncommutative parameter $\theta$, one
should not expect any significant departure in the interference pattern from that of the usual Aharonov-Bohm effect
in the normal plane.

In the following section we will analyze the equation of motion of the electron in the
aforementioned magnetic field, retaining modifications due to non-commutativity up to first order in the parameter
$\theta$.

\section{The Schr\"{o}dinger-Pauli Equation}

In this section we write down the explicit form of the Schr\"odinger-Pauli equation in the presence of the
gauge field (\ref{GNC}). Let us consider the Hamiltonian
\begin{equation}\label{52}
    H= \frac{1}{2} \left(\hat{\mathbf{p}} - e \mathbf{A}^{\nc}\right)^{2} = \frac{1}{2}\left(-\imath \BS{\nabla} - {e}\, \mathbf{A}^{\nc} \right)^2\,,
\end{equation}
corresponding to a non-relativistic spinorial particle ($m=1$). The Hamiltonian $H$ is invariant under a rotation by an angle $\gamma$ around the $z$-axis under which the vector potential transforms as
\begin{equation}\label{53}
    \mathbf{A}^{(NC)} \rightarrow U(\gamma) \mathbf{A}^{(NC)} U^{\dagger}(\gamma)
\end{equation}
with $U(\gamma):=e^{\imath \gamma J}$ defined in \eqref{45}. As a result, $H$ commutes with $J:=L+\sigma_3/2$ and it leaves invariant the subspaces of the form
\begin{equation}\label{60}
    \mathcal{H}_l:=\left\{ \left(
\begin{array}{c}
 e^{i l \phi }\, \psi_l (r) \\
 e^{i (l+1) \phi }\, \chi_l (r)
\end{array}
\right)\, : \,  \psi_l (r),\chi_l (r)\in \mathbf{L}_2\left(\mathbb{R}^+; r\, dr\right)\right\}\,,
\end{equation}
for all $l\in \mathbb{Z}$. The eigenvalue equation for $H$ in this subspace gives rise to a system of coupled differential equations  which, in the leading order in the perturbation parameter $\theta$,  gives a good description of our system at large $r$.  As we will show, it is possible to get an exact expression for this first order correction.

For small $\theta$, eigenvalue equations for $H$ reduces to
\begin{align}\label{64}
      & -\psi_l''(r)-\frac{1}{r}\,\psi_l '(r)+
    \left\{\frac{(l-\beta)^2 }{r^2} -E
    \right\} \psi_l(r) =- \theta \beta \left\{- \frac{1}{r^2}\, \chi_l'(r)+ \left(\frac{l+1-\beta}{r^3}\right)\chi_l (r)\right\}
      + O\left(\theta^2\right)\,,\nonumber\\
      & -\chi_l''(r)-\frac{1}{r}\, \chi_l '(r)+
      \left\{\frac{(l+1-\beta)^2}{r^2}
      -E\right\}\chi_l (r)  =-\theta \beta \left\{
      \frac{1}{r^2}\, \psi_l '(r)+\frac{l-\beta}{r^3}\, \psi_l (r)\right\} + O\left(\theta^2\right)\,.
\end{align}
with $ \beta:= e \alpha$. Let us note that, although for small $r$ the coefficients in the eigenvalue equation are all  regular,
the small-$\theta$ expansion introduces singular terms  (at $r=0$) which we here treat as perturbations on the solutions of the usual Aharonov-Bohm problem.

It is worthwhile to point out that the first order perturbed Hamiltonian can be constructed through successive unitary
transformations from that of the standard Aharonov-Bohm Hamiltonian, inheriting therefore the spectrum and topological properties of the last one. The noncommutative effects in the present approach includes anisotropic ones that could --in  principle-- be measured in experiments.

Indeed, up to first order in $\theta$, the Hamiltonian in Eq.\ (\ref{52}) can be written as

\bb
{ H} = -\frac{1}{2}\left[\left\{\left(\frac{\partial}{\partial r} +\frac{1}{2r}\right)^2 + \frac{1}
{4r^2} + \frac{1}{r^2} \left(\frac{\partial}{\partial \phi} -i \beta \right)^2 \right\}\BS 1_2+ \frac{i\,\beta \theta}{r^2} \left\{\sigma_{\phi} \frac{\partial}{\partial r} +
\frac{\sigma_{r}}{r}
\left(\frac{\partial}{\partial \phi} -i \beta \right) \right\}\right], \label{ha1}
\ee
and using the identities
\begin{equation}
r^{-1/2} \frac{\partial}{\partial r} [r^{1/2}~~]  = \frac{\partial}{\partial r} + \frac{1}{2r},
\qquad e^{\imath \beta \phi} \frac{\partial}{\partial  \phi} [e^{-\imath \beta \phi}~~] = \frac{\partial}
{\partial \phi} -\imath \beta,
\end{equation}
it can also be factorized as
\eqb
{ H} &=& -\frac{1}{2} r^{-1/2} e^{\imath \beta \phi} \left[ \left\{\frac{\partial^2}{\partial r^2} +
\frac{1}{4r^2} + \frac{1}{r^2} \frac{\partial^2}{\partial \phi^2}\right\}\BS 1_2 + \frac{\imath \beta \theta}
{r^2} \left\{\sigma_\phi \left( \frac{\partial}{\partial r} - \frac{1}{2r}\right) +
\sigma_r\,\frac{1}{r} \frac{\partial}{\partial \phi}\right\} \right]  r^{1/2} e^{-\imath \beta \phi},
\nonumber\\
&\equiv&\bigg( r^{-1/2} e^{\imath \beta \phi}\bigg)\, {\tilde H}\,
\bigg(r^{1/2} e^{-\imath \beta \, \phi}\bigg).
\label{uni}
\eqf
Here  ${\tilde H}$ is
\bb
{\tilde H} ={\tilde  H}_0 - \theta\frac{\imath \beta }{2r^2} \left[\sigma_\phi \left( \frac
{\partial}{\partial r} - \frac{1}{2r}\right) +
\sigma_r\, \frac{1}{r} \frac{\partial}{\partial \phi}\right], \label{til1}
\ee
with
\bb
{\tilde  H}_0= -\frac{1}{2}\left[\frac{\partial^2}{\partial r^2} +\frac{1}{4r^2} +\frac{1}
{r^2} \frac{\partial^2}{\partial \phi^2}  \right] \BS 1_2. \label{til2}
\ee

The Hamiltonian  $\tilde {H}$ can be factorized again by using  the following identity
\bb
\left[ {\tilde H}_0, \frac{i\,\beta \,\theta}{2 r} \sigma_\phi \right] = \frac{i\,\beta \,\theta}
{2 \,  r^2} \left(\sigma_{\phi} \left[\frac{\partial}{\partial r} - \frac{1}{2r}\right] +\sigma_{r}\, \frac{1}{r} \frac{\partial}
{\partial \phi} \right);
\ee
the relation between ${\tilde H}$ and ${\tilde H}_0$ becomes now
\eqb
{\tilde H} &=&{\tilde  H}_0 -\left[ {\tilde H}_0, \frac{i\,\beta \,\theta}{2 r} \sigma_\phi \right],
\nonumber
\\
&=& \left( 1 + \frac{i\,\beta \,\theta}{2 r} \sigma_{\phi} \right) {\tilde H}_0
\left( 1 - \frac{i\,\beta \,\theta}{2 r} \sigma_{\phi} \right) \nonumber\\
&= & { U}  {\tilde H}_0 { U}^{\dagger}, \label{apro1}
\eqf
up to first order in $\theta$.

Observe that the multiplicative unitary operator $U=\left(1 +\frac{i\,\beta\,\theta}{2r}\,\sigma_{\phi}\right)$ (which
encloses all the dependence on $\theta$) factorizes out the nontrivial matrix dependence due to spin, leaving ${\tilde H}_0$  which is diagonal. Moreover, it commutes with the factor $r^{1/2}e^{-ie\alpha\phi}$ in the similarity transformation in (\ref{uni}) so that up to first order in $\theta$, the Hamiltonian $\hat{H}$ describing the modified Aharonov-Bohm effect can be written as
\begin{equation}
\label{factor2}
{H} = U\,\bigg[ \big( r^{-1/2} e^{\imath \beta \phi}\big)\,\tilde{H}_0\,\big( r^{1/2} e^{-\imath \beta
 \phi}\big)\bigg]\,U^\dag.
\end{equation}
where the term in parentheses is just the standard Aharonov-Bohm Hamiltonian, which acts as a diagonal
operator on the spinor components.

Achieved this result, it is easy to relate the eigenfunctions $\psi_E(r,\phi)$ of the perturbed Hamiltonian in Eq.\ (\ref{ha1}),
\bb
{H}\, \psi_E(r,\phi) =E\,\psi_E(r,\phi), \label{prob}
\ee
with the corresponding eigenfunction of the usual Aharonov-Bohm scalar particle on the usual commutative plane,
through the above discussed unitary transformation. In particular, notice that the spectrum of the Hamiltonian in the
noncommutative case is the same as that of the standard Aharonov-Bohm effect.

Indeed, if
\bb
{\tilde H}_0\, \tilde{\chi}_E^{(0)} (r,\phi)= E\,\tilde{\chi}_E^{(0)}(r,\phi) \label{ei}
\ee
then, from Eq.\ (\ref{factor2}), we get
\bb
\label{final}
\psi_E (r,\phi) = {r^{-1/2}}\, e^{\imath \beta \phi}\, { U} \tilde{\chi}_E^{(0)} (r,\phi).
\ee
Notice that the separability of variables in each component of the solution is ensured by the previously discussed
rotational symmetry of the Hamiltonian.

It is straightforward to find  solutions of (\ref{ei}): each component of the spinor $\tilde{\chi}_E^{(0)}(r,\phi)$ (which are not
coupled) can be expressed as
\[
\tilde{\chi}_j (r,\phi) =  e^{i \nu \phi} {\chi}_j^{\nu} (r),\qquad j=1,2\,,
\]
where the parameters $\nu$ are not necessarily integers, since it is the function $\psi_E (r,\phi)$ in Eq.\ (\ref{final}) which
must be single-valued. The radial functions ${\chi}_j^{\nu}(r)$ satisfy the equation
\bb
 -\frac{1}{2}\left[\frac{d^2}{dr^2} +\frac{1-4\nu^2}{4r^2} \right] { \chi}_j^{\nu} (r) = E\,
{ \chi}_j^{\nu} (r), \label{esta}
 \ee
whose solutions are expressed in terms of Bessel Functions as
\bb
{\chi}_j^{\nu}(r)= \sqrt{r} \left[
A_j  J_{|\nu|}(k r) + B_j Y_{|\nu|} (k r)\right],
\ee
where $k=\sqrt{2E}$ and $A_j,B_j$ $(j\in\{1,2\})$ are integration constants which must be determined according to the
boundary condition the function $\psi_E (r,\phi)$ must satisfy.

Comparison with Eq.\ (\ref{60}) shows that the parameter $\nu$ must be chosen so as
\begin{equation}\label{autofuncion}
    \psi_E^\ell (r,\phi)= \left(1 +\imath\, \frac{\beta\,\theta}{2r}\,\sigma_{\phi}\right)
    \left(
\begin{array}{c}
 e^{i \ell \phi } \left[A_1^\ell  J_{|l-\beta|}(k r) + B_1^\ell Y_{|\ell-\beta|} (k r)\right] \\ \\
 e^{i (\ell+1) \phi }\, \left[A_2^\ell  J_{|\ell+1-\beta|}(k r) + B_2^\ell Y_{|\ell+1-\beta|} (k r)\right]
\end{array}
\right)\,.
\end{equation}

The coefficients in Eq.\ (\ref{autofuncion}) must be determined by imposing suitable boundary conditions. In particular,
this first order correction in perturbation theory must be square-integrable in a neighborhood of the origin. It can be
straightforwardly seen that this condition requires that $B_j^\ell=0\,, j=1,2\,, \forall \ell \in \mathbf{Z}$.

Employing the recurrence relation for Bessel functions,
\begin{equation}\label{recurrence}
    Z_{n+1}(x)+Z_{n-1}(x)=\frac{2 n}{x}\, Z_{n}(x)\,,
\end{equation}
the solution in Eq.\ (\ref{autofuncion}) can also be written as
\begin{equation}\label{autofuncion-J}
\psi_E^\ell (r,\phi)=
\left(
\begin{array}{c}
    {e^{i \ell \phi } }
    \left\{{A_1^\ell}
   J_{|\ell-\beta |}(k r)+\frac{\theta\beta k   \,{A_2^\ell} }{{4 |\ell-\beta +1|}}
   \left[J_{|\ell-\beta +1|-1}(k r)+J_{|\ell-\beta +1|+1}(k
   r)\right]\right\} \\ \\e^{i (\ell+1) \phi }\left\{
   {A_2^\ell}
   J_{|\ell-\beta +1|}(k r)-
 \frac{\theta \beta k\, {A_1^\ell}
   }{4 |\ell-\beta |}\left[J_{|\ell-\beta |-1}(k r)+J_{|\ell-\beta |+1}(k r)\right]\right\}
\end{array}
\right)\,.
\end{equation}

The general solution of the Schr\"{o}dinger-Pauli equation in the noncommutative plane, up to first order in $\theta$, is
then constructed as the combination\footnote{Strictly speaking, we should consider also linearly independent (order $\theta$) square-integrable solutions in the $\ell=-1,0,1$ invariant subspaces, which have a nonregular behavior at the origin and are related to the existence of nontrivial self adjoint extensions of the Aharonov-Bohm Hamiltonian \cite{FP} (See Appendix \ref{Y}). For simplicity, we impose a regularity condition at the origin on the solutions of the Aharonov-Bohm problem from which we construct the eigenfunctions we consider in the following.}
\begin{equation}
\label{exact}
\psi_E (r,\phi) = \sum_{\ell=-\infty}^\infty \psi_E^\ell (r,\phi)\,,
\end{equation}
with $\psi_E^\ell (r,\phi)$ given in Eq.\ (\ref{autofuncion}) or (\ref{autofuncion-J}).

With this solutions, our aim is now to evaluate the scattering amplitude, which could be of interest
for an experimental test of this kind of systems. This will be done in the next section.

\section{Scattering Amplitude}

We are now interested in a situation in which an incident particle reaches the center at $r=0$ and is scattered out by the
Aharovnov-Bohm flux in the noncommutative plane we are considering. In this case, the wave function constructed in
Eq.\ (\ref{exact}) equals the sum of the plane wave and a spherical outgoing wave.

The asymptotic form of the solution (\ref{exact}), for $kr>>1$ turn out to be
\begin{equation}
\label{exactexpand}
\psi_E(r,\phi) \approx \left(
\begin{array}{c}\displaystyle
\sum_{\ell =-\infty}^{\infty} e^{i\ell\phi}\,A_1^\ell\,\sqrt{\frac{2}{\pi k r}}
\cos\left(kr - \frac{\pi}{2}|\ell -\beta|-\pi/4\right)
\\
\displaystyle
\sum_{\ell =-\infty}^{\infty} e^{i(\ell+1)\phi}\,A_2^\ell\,\sqrt{\frac{2}{\pi k r}}
\cos\left(kr - \frac{\pi}{2}|\ell -\beta+1|-\pi/4\right)
\end{array}
\right)
\end{equation}
and we observe that there are no terms proportional to $\theta$. This is completely different than in conventional quantum mechanics where the differential cross section \cite{weab} may depend on theta in the case of small angle.

Therefore, the differential cross section for our problem is the same as in the Aharonov-Bohm
effect for spin $1/2$ particles. As a check, consider a polarized beam with $A_2^\ell =0$.
It is direct to check that previous expression has the shape
$$
\psi_E(r,\phi) = \left(
\begin{array}{c}
e^{ikr\cos(\phi)}
\\
0
\end{array}\right) +
\left(
 \begin{array}{c}\displaystyle
F_1(\phi)\,\frac{e^{ikr}}{\sqrt{ir}}
\\
0
\end{array}\right)
$$
by choosing
\begin{equation}
A_1^\ell =e^{\frac{i\pi}{2}(2\ell - |\ell -\beta|)},
\end{equation}
and then
\begin{equation}
\label{fs}
F_1 =\frac{1}{\sqrt{2k\pi}}\sum_{\ell=-\infty}^\infty e^{i\ell\phi}\left(e^{i\pi(\ell - |\ell - \beta|)}-1\right),
\end{equation}
which  is the standard result  for the Aharonov-Bohm effect (see for example \cite{weab}).
Therefore, the differential cross section, for a beam polarized parallel to the solenoid direction, the
cross section   turns out to be
\begin{equation}
\label{dcs}
\frac{d\sigma}{d	\phi}  = \frac{\sin^2(\pi\beta)}{2k\pi\sin^2(\phi/2)}.
\end{equation}

Opposite  polarization, namely $A_1^\ell =0$, gives same result, as it is expected from the general result
\cite{hagen,schmidt}
\begin{equation}
\label{genres}
\frac{d\sigma}{d	\phi} =\bigg(1-(\hat{{\bf n} }\times \hat{{\bf z}})^2\,\sin^2(\phi/2)\bigg)\left(\frac{d\sigma}{d	 \phi}\right)_{\mbox{\small{unpol}}},
\end{equation}
where $\hat{\bf n}$ is the polarization direction and $\hat{\bf z}$, the direction defined by the solenoid.

\medskip

Anyway, a comment is in order. As previously stated, we are employing perturbation theory, which introduces potentials which are singular at the origin, even though the noncommutative of this plane smears out the field intensity as discussed in Section \ref{NC-plane}, leading to smooth coefficients in the Hamiltonian.
Unfortunately, we were not able to solve the equations with the exact (regular) potential, and rather make use of perturbation theory. But one should keep in mind that the exact solution could give different phase shifts and affect the expression of the cross-section.

\section{Conclusions}

In this paper we have studied the Aharonov-Bohm effect with  spin-noncommutative, namely, for the case when the noncommutative of coordinates involves spin. Although the Aharonov-Bohm effect, in this case, retains some important properties of the conventional case, such as topological properties, spin induces a strong anisotropy which is a completely different from the conventional Aharonov-Bohm effect. From the analysis of the scattering problem one sees that the anisotropy is an effect that occurs close to $\theta \sim \sqrt {r} $ so that long-range scattering effects may not be affected.

However, this anisotropy opens also the interesting possibility of studying other
systems that behave very similarly to the problem studied here. Indeed, in the case of
cold atoms with nonzero total spin using suitable magnetic traps it is possible to
confine the atoms to a plane and, therefore, such as in the case of an anyon  gas,
one could have a gas of cold atoms where each atom would have an attached magnetic  flux
as in the conventional anyon gas \cite{wilczek}. The experiments measuring spin effects of cold atoms in gases have so far been done only for the case of total spin 3. More accurate measurements in other systems are likely to  come in near future where predictions of spin-noncommutative can possibly be checked.

\vspace{0.3 cm}

\noindent{\textbf{Acknowledgements}}:  This  work was supported  in part by US DOE Grant number DE-FG 02-91ER40685, by FONDECYT-Chile grant-1095106,  1100777, Dicyt (USACH), and CONICET (PIP 01787), ANPCyT (PICT 00909) and by UNLP (Proy.~11/X492), Argentina. M.N. also acknowledge support from Universidad Nacional de La Plata, Argentina.

\appendix

\section{Solutions in the critical subspace}\label{Y}
In the \emph{critical} ${\ell} = -1,0,1$ invariant subspaces, we can find linearly independent $O(\theta)$ solutions (eigenfunctions of the Aharonov-Bohm Hamiltonian), whose existence is related to the existence of nontrivial self-adjoint extensions of this Hamiltonian \cite{FP}. Indeed, the functions
\begin{equation}\label{critical}
    \zeta^{-1}_E=\left(
                  \begin{array}{c}
                    0    \\
                    \theta B_2^{-1} \, Y_\beta(k r) \\
                  \end{array}
                \right)\,,
                \quad
     \zeta^{0}_E=\left(
                  \begin{array}{c}
                   \theta B_1^{0} \, Y_\beta(k r)    \\
                    \theta B_2^{0}e^{i  \phi } \, Y_{1-\beta}(k r) \\
                  \end{array}
                \right)\,,
                \quad
                 \zeta^{1}_E=\left(
                  \begin{array}{c}
                     \theta B_2^{-1}e^{i  \phi } \, Y_{1-\beta}(k r)    \\
                   0 \\
                  \end{array}
                \right)\,,
\end{equation}
are square-integrable near the origin (with respect to the measure $r\, dr\, d\phi$) and satisfy (up to order $\theta$) the eigenvalue equation (\ref{prob}) with $2 E=k^{2}$. Then, in these critical subspaces, one should take those linear combinations of functions in Eqs.\ (\ref{autofuncion-J}) and (\ref{critical}) which belong to the domain of the selected Hamiltonian self-adjoint extension. They are of the form
\begin{equation}\label{lmenos1}
    \begin{array}{c}
       \psi_E^{-1} (r,\phi)=
\left(
\begin{array}{c}
    {e^{- i  \phi } }
    \left\{{A_1^{-1}}
   J_{1+\beta }(k r)+\frac{\theta\beta k   \,{A_2^{-1}} }{{4 \beta }}
   \left[J_{\beta -1}(k r)+J_{\beta +1}(k
   r)\right]\right\} \\ \\\left\{    {A_2^{-1}}
   J_{\beta}(k r)-  \frac{\theta \beta k\, {A_1^{-1}}
   }{4 (1+\beta )}\left[J_{\beta }(k r)+J_{2+\beta }(k r)\right]+
   \theta B_2^{-1} \, Y_\beta(k r)\right\}
\end{array}
\right) \simeq
\\ \\
     \simeq
\left(
  \begin{array}{c}
    r^{\beta -1} \left[\theta\,\frac{2^{-\beta -1}   A_2^{-1} k^{\beta } }{\Gamma
   (\beta )}+ O\left( r^2 \right) \right]
   \\   \\
    r^{-\beta } \left[-\theta\, \frac{2^{\beta } k^{-\beta } \theta  B_2^{-1} \Gamma
   (\beta )}{\pi }+O\left(r^2\right)\right]
   + O\left(r^\beta\right)
  \end{array}
\right)\,,
    \end{array}
\end{equation}
\begin{equation}\label{l0}
    \begin{array}{c}
        \psi_E^0 (r,\phi)=
\left(
\begin{array}{c}
    \left\{{A_1^0}
   J_{\beta }(k r)+\frac{\theta\beta k   \,{A_2^0} }{{4 (1-\beta)}}
   \left[J_{-\beta}(k r)+J_{\beta+1}(k
   r)\right]+\theta B_1^0 Y_\beta(k r)\right\} \\ \\e^{i  \phi }\left\{
   {A_2^0}
   J_{1-\beta}(k r)-
 \frac{\theta \beta k\, {A_1^0}
   }{4 \beta }\left[J_{\beta -1}(k r)+J_{\beta +1}(k r)\right]+
   \theta B_2^{0} Y_{1-\beta}(k r)\right\}
\end{array}
\right)\,.\simeq
\\ \\
\left(
  \begin{array}{c}
    r^{-\beta }\left[\theta 2^{\beta }  k^{-\beta } \left(\frac{k \beta
   A_2^0}{(4-4 \beta ) \Gamma (1-\beta )}-\frac{B_1^0 \Gamma (\beta
   )}{\pi }\right)+O\left(r^2\right) \right] +O\left(r^\beta\right)
   \\   \\
   r^{1-\beta }\left[\frac{2^{\beta -1} k^{1-\beta }  (\pi  A_2^0+\theta
   B_2^0 \cos (\pi  \beta ) \Gamma (2-\beta ) \Gamma (\beta -1))}{\pi
   \Gamma (2-\beta )}+O\left(r^2\right)\right]
   +r^{\beta -1} \left(-\frac{2^{-\beta -1} k^{\beta -1} \theta  (k \pi
   A_1^0+4 B_2^0 \Gamma (1-\beta ) \Gamma (\beta ))}{\pi  \Gamma
   (\beta )}+O\left(r^2\right)\right)
  \end{array}
\right)\,,
    \end{array}
\end{equation}
\begin{equation}\label{l1}
    \begin{array}{c}
       \psi_E^1 (r,\phi)=
\left(
\begin{array}{c}
    {e^{i  \phi } }
    \left\{{A_1^1}
   J_{1-\beta }(k r)+\frac{\theta\beta k   \,{A_2^1} }{{4 (2-\beta)}}
   \left[J_{1-\beta}(k r)+J_{3-\beta }(k   r)\right]\right\} \\ \\e^{i 2 \phi }\left\{
   {A_2^1}    J_{2-\beta}(k r)-  \frac{\theta \beta k\, {A_1^1}
   }{4 (1-\beta )}\left[J_{-\beta }(k r)+J_{-\beta }(k r)\right]\right\}
\end{array}
\right) \simeq
\\ \\
     \simeq
\left(
  \begin{array}{c}
  r^{\beta -1}\left[ -\theta \frac{2^{1-\beta }   B_1^1 k^{\beta -1}  \Gamma
   (1-\beta )}{\pi }+O\left( r^2 \right) \right]+O\left( r^{1-\beta} \right)
   \\   \\
  r^{-\beta } \left[ \theta\frac{ 2^{\beta -2} \beta    A_1^1 k^{1-\beta } }{(\beta -1) \Gamma (1-\beta )}+O\left( r^2 \right)\right]
  \end{array}
\right)\,,
    \end{array}
\end{equation}
where the coefficients are to be fixed by imposing the conditions the functions in the domain of the operator satisfy near the origin.

As previously mentioned, in the present article we discard such contributions and impose just regularity at the origin on the solutions of the Aharonov-Bohm problem from which we construct the eigenfunctions of the present one. But it is worth to mention that these additional contributions open the possibility of a spin-flip in the critical subspaces. This will be considered elsewhere.

\end{document}